\documentclass[prd,letterpaper,tightenlines,nofootinbib,preprint,floatfix,showpacs]{revtex4}
\newcommand{\z}[1]{\left({#1}\right)}

\def\d{\mathrm{d}}
\def\bvec#1{{\rm\bf #1}}

\usepackage{graphicx}
\usepackage{epsfig}
\begin{document}

\title{A new family of exact and rotating solutions of fireball
hydrodynamics\footnote{Dedicated to T. Kodama on the occasion of his 70th birthday.}}
\author{T. Cs{\"o}rg{\H o}}\affiliation{Wigner RCP, H - 1525 Budapest 114, P.O.Box 49, Hungary}
\author{M. I. Nagy}\affiliation{Institute of Physics, E\"otv\"os University, H-1117 Budapest, P\'azm\'any P. s. 1/A, Hungary,}
\date{December 30, 2013}

\begin{abstract}
A new class of analytic, exact, rotating, self-similar and surprisingly simple solutions of non-relativistic
hydrodynamics are presented for a three-dimensionally expanding, spheroidally symmetric fireball.
These results generalize earlier, non-rotating solutions for ellipsoidally symmetric fireballs with directional,
three-dimensional  Hubble flows. The solutions are presented for a general class of equations of state that
includes the lattice QCD equations of state and may feature inhomogeneous temperature and corresponding density
profiles.
\end{abstract}

\pacs{24.10.Nz,47.15.K}
\maketitle

\section{Introduction} 
Heavy ion collisions with finite impact parameters create systems where the net angular momentum is not
zero in the initial state, and so (due to angular momentum conservation) this creates a rotating and expanding
fireball. Recently, rotating numerical solutions of hydrodynamics became a subject of intensive theoretical
interest and new observables are designed~\cite{Csernai:2013uda,Csernai:2013vda, Csernai:2013add}
to search for the effects of rotation in the experimental data.

Our goal is to generalize earlier exact analytic solutions of fireball hydrodynamics to rotating flows.
There are many already known, important, analytic and self-similar exact solutions of fireball hydrodynamics. 
The first solutions in this class,
the Zim\'anyi-Bondorf-Garpman (ZBG) solution~\cite{Bondorf:1978kz} inspired the Cs\"org\H{o}-Csizmadia-Luk\'acs
(CsCsL) solution for a 3d expanding Gaussian fireball with Hubble flow and with a spatially homogeneous temperature
profile~\cite{Csizmadia:1998ef} which was the first solution of fireball hydrodynamics with a Gaussian density profile
and simple scaling expressions for the mass dependence of the radial flow and the HBT radii.
This solution has later been generalized to inhomogeneous temperature profiles with spherical symmetry
in ref.~\cite{Csorgo:1998yk}, subsequently the spherical symmetry was generalized to a 
three-dimensional, ellipsoidally symmetric expansion in refs.~\cite{Akkelin:2000ex,Csorgo:2001ru},
and to expansions with rather arbitrary equations of state (corresponding to temperature dependent speed of sound) in ref.~\cite{Csorgo:2001xm}.

As is well known, the first exact  solutions of relativistic hydrodynamics were described by Landau and collaborators,
who improved on Fermi's statistical model, observing that if thermodynamical laws determine the production cross-sections
of strongly interacting particles in high energy (cosmic ray induced) reactions, then the momentum distributions
of these particles needs to be calculated from the theory of locally thermalized and relativistically expanding fluids~\cite{Landau:1953gs,Belenkij:1956cd}. 
Unfortunately, the Landau solutions were given in a complicated, implicit form.
In the ultra-relativistic, infinite bombarding energy limit, a self-similar and explicit scaling solution was found by R. C. Hwa~\cite{Hwa:1974gn}
and later, independently by Bjorken~\cite{Bjorken:1982qr}, who, with the help of this solution, figured out how to estimate the initial
energy density of high energy proton-proton or heavy ion reactions from  experimental data. His paper on relativistic hydrodynamics
became a fundamental contribution the field of relativistic heavy ion collisions, although the resulting boost-invariant 
rapidity distribution is not seen even at the present, $\sqrt{s} = 7$ TeV LHC energies.

The CsCsL solution has also been extended to the domain of relativistic
kinematics: first a one-dimensional, longitudinally expanding explicit relativistic solution has been found in ref.~\cite{Csorgo:2002ki},
then this class was generalized to relativistic expansions with three dimensional, axially/cylindrically symmetric
expansions in refs.~\cite{Csorgo:2002bi,Csorgo:2003rt} and soon also to the case of three dimensional, relativistic
expansions with a realistic, ellipsoidal symmetry of the temperature and the density profile~\cite{Csorgo:2003ry}.
Although these relativistic solutions match the non-relativistic solutions in the limit of non-relativistic velocities,
they all had shortcomings: the lack of the relativistic acceleration, and the presence of the spherically symmetric,
direction independent Hubble flow profile in each of them.

The search started for accelerating and relativistic solutions of hydrodynamics, that may lead to finite rapidity distributions.
This resulted in the discovery of several non-Hubble type simple and explicit, exact solutions of relativistic hydrodynamics, as
reported first in refs.~\cite{Csorgo:2008pe,Csorgo:2006ax} and described in full detail in ref.~\cite{Nagy:2007xn},
these are the so-called NCC solutions. 

The CsCsL solution of ref.~\cite{Csizmadia:1998ef} was special  in another, rather unusual manner too, being, quite unexpectedly,
a simultaneous solution of two, generally quite different dynamical equations: the collisionless Boltzmann equation as well as
the system of equations of hydrodynamics, for both in the non-relativistic kinematic region. Some of the new,
accelerating relativistic hydrodynamical solutions of refs.~\cite{Csorgo:2008pe,Csorgo:2006ax,Nagy:2007xn} were also
solutions of the collisionless Boltzmann equations. The systematic search for such simultaneous, so called ``collisionless''
solutions started and one of the important and interesting outcome of these studies were reported in
refs.~\cite{Nagy:2009eq,nagymarciPhD}: not only that another rich class of accelerating relativistic hydrodynamics was
found but also the first exact, explicit and {\it rotating} solution of relativistic hydrodynamics was described.

The effort to solve the relativistic hydrodynamical equations for cases relevant
to heavy-ion physics phenomenology is an ongoing research theme, that undergoes an interesting revival and resulted
very recently in decent, new exact solutions. 
The Landau hydrodynamics has also been revisited and the evaluation of the rapidity distribution and particle spectra
has been improved in ref.~\cite{Wong:2008ex}. It also served as a basis for more general (albeit approximate)
hydrodynamical solutions that enabled the authors to study transverse dynamics in the framework of the
Landau model~\cite{Jiang:2013rm,Zetenyi:2010jp,Tamosiunas:2011qg}. A unified description of the famous
Hwa-Bjorken and the Landau-Khalatnikov solutions was reported in ref.~\cite{Bialas:2007iu} 
that has been extended to a general treatment of 1+1 dimensional relativistic flows in ref.~\cite{Beuf:2008vd,Peschanski:2010cs}.
Recently, multi-dimensional solutions of relativistic hydrodynamics were reported, that include transverse
dynamics in an advanced manner~\cite{Liao:2009zg,Lin:2009kv,Gubser:2010ze}.

In this manuscript, we describe rotating solutions of non-relativistic hydrodynamics that feature spheroidal
(ie. rotationally symmetric ellipsoidal) density profiles~\cite{csorgo:zimanyi2012}. (A very special case of
these solutions was briefly considered in ref.~\cite{nagymarciPhD}.) For spatially homogeneous temperature
profiles, we obtain solutions that can incorporate fairly general equations of state, based on a non-relativistic
analogy of the relativistic, irrotational and non-acccelerating solutions that were given for a broad class of
equations of state, including the lattice QCD equations of state~\cite{Csanad:2012hr}.
In the limiting case of no rotation, the rotating solutions reported here reproduce 
already known, irrotational solutions of non-relativistic fireball hydrodynamics.

\section{Notation and basic equations}
We consider a non-relativistic fluid, a fireball with an adiabatic expansion, assuming that the bulk and shear
viscosity as well as the heat conductivity is negligible and the expanding fluid is locally thermalized. Time
is denoted by $t$, the spatial coordinate by $\bvec r= (r_x, r_y, r_z)$, the mass of the particles is $m$ and
the hydrodynamical fields such as the number density $n$, the pressure $p$, the energy density $\varepsilon$
and the temperature $T$ generally depend on both time and space: $n \equiv n(t,\bvec r)$, $p \equiv p(t,\bvec r)$, 
$\varepsilon \equiv \varepsilon(t,\bvec r)$ and $T\equiv T(t,\bvec r)$. In the non-relativistic case, the
dynamical equations that govern the expansion of such fireballs are the continuity, Euler and energy
equations, corresponding to the conservation of matter (charge), momentum and energy. These equations read as follows:
\begin{eqnarray}
{\partial_t n} +\bvec \nabla (n \bvec v) & = &0,  \label{e:cont} \\
(\partial_t  +\bvec v \bvec\nabla) \bvec v &=&- \frac{\bvec \nabla p}{m
n}, \label{e:Euler} \\
{\partial_t \varepsilon } + \bvec \nabla  (\varepsilon \bvec v)  + p \bvec \nabla \bvec v & = & 0,  \label{e:energy}
\end{eqnarray}
and they are closed by the so-called equations of state, that characterize the local thermodynamical properties 
of the flowing matter. Following ref.~\cite{Csorgo:2001xm}, we assume that the ratio of energy density to the pressure 
is an arbitrary, temperature dependent function:
\begin{eqnarray}
p & = & n T, \label{e:pnt}  \\
\varepsilon &=& \kappa(T) p. \label{e:ekp} 
\end{eqnarray}
Fundamental thermodynamical relations, irrespectively of the
equations of state, read as 
\begin{eqnarray}
 \varepsilon + p & = & \mu n + T \sigma , \\ \label{e:tdk}
 \d\varepsilon & = &  \mu \d n + T \d\sigma ,\\ \label{e:de}
 \d p &  = & n \d \mu + \sigma \d T, \label{e:dp}
\end{eqnarray}
where $\mu $ stands for the chemical potential and $\sigma$ denotes the entropy density.
These fundamental relations, together with the local conservation of energy as
expressed by eq.~(\ref{e:energy}), imply the local conservation of the entropy
density,
that corresponds to the adiabatic nature of non-dissipative flows:
\begin{equation}
 {\partial_t \sigma} +\bvec \nabla (\sigma \bvec v)  = 0.  \label{e:entropy} \\
\end{equation}

We consider another scenario as well, when there is no conserved charge or particle number density $n$ present.
This scenario was recently considered in detail in the relativistic kinematic domain in ref.~\cite{Csanad:2012hr}.
In this scenario, the equation of state is as follows:
\begin{eqnarray}
 \varepsilon + p & =& T \sigma,\label{e:eos-non}\\
 \varepsilon & = & \kappa(T) p. \label{e:eos-non2}
\end{eqnarray}
Note that the energy equation, eq.~(\ref{e:energy}) can be utilized with this
equation of state without any conceptual difficulty,
however, the Euler equation needs to be modified in the $n \to 0$ limit. This
can be done by starting with the relativistic form of the Euler equation, and
then considering the $v \ll c$ limit.
Thus in the considered, $n \to 0$ scenario, the Euler equation becomes modified as follows:
\begin{equation}
 ( \varepsilon + p) (\partial_t  +\bvec v \bvec\nabla) \bvec v = - \bvec \nabla p, \\ \label{e:Euler-non}
\end{equation}
which is an interesting form of the non-relativistic Euler equation for a fluid where there is no conserved
charge. In this case, a lattice QCD equation of state can be used to follow a non-relativistic fireball explosion,
similarly to the recently found relativistic hydrodynamical solutions described in ref.~\cite{Csanad:2012hr}.

We emphasize that in fact we find two different families of solutions of non-relativistic hydrodynamics. 
The first family, {\it Case 1} corresponds to the solution of the conventional
form of the equations of
non-relativistic hydrodynamics, as specified by eqs.~(\ref{e:cont},\ref{e:Euler},\ref{e:energy}) together with the equations of state
as given by eqs. (\ref{e:pnt},\ref{e:ekp}). The second family, {\it Case 2} corresponds to the
solutions of the energy equation, eq.~(\ref{e:energy}) (or, alternatively,
eq.~(\ref{e:entropy}), the entropy conservation equation) and
the Euler equation as given by (\ref{e:Euler-non}), with the equation of state given by eqs.~(\ref{e:eos-non},\ref{e:eos-non2}).
Both in \emph{Case 1} and \emph{Case 2} one can use general $\kappa \equiv \kappa(T)$ functions in the equation of state. In \emph{Case 2},
this allows one to include the lattice QCD equations of state, which is characterized by lack of conserved charge in the baryon-free, 
$\mu = 0$ region.

A simplification of the equations is achieved if one rewrites the equations for the temperature $T$.
One can verify that in \emph{Case 1}, by substituting the equation of state into the Euler, energy,
and continuity equations, we obtain 5 partial differential equations for 5 unknown fields
(temperature $T$, conserved number density $n$, and the three components of the velocity
field $\bvec v$), as follows:
\begin{eqnarray}
{\partial_t n} +\bvec \nabla (n \bvec v) & = &0,  \label{e:cont-class1} \\
 \left[\frac{\d }{\d T}(\kappa T) \right]\z{\partial_t + \bvec v \bvec \nabla} T + T\bvec \nabla \bvec v & = & 0,  \label{e:T-class1} \\
n m (\partial_t  +\bvec v \bvec\nabla) \bvec v &=& - \bvec \nabla p  ,\label{e:Euler-class1} \\
p & = &nT .
\end{eqnarray}
In this case, the entropy density $\sigma$ is not an unavoidable unknown field, but can readily be calculated if necessary from the
fundamental thermodynamical relations and from the equation of state.

For \emph{Case 2}, a similar system of equations is obtained for the 5 unknown fields
(entropy density $\sigma$, temperature $T$, and $\bvec v$). These equations read as follows:
\begin{eqnarray}
 {\partial_t \sigma} +\bvec \nabla \sigma \bvec v & = &0,  \label{e:entropy-class2} \\
 \frac{1+\kappa}{T}\left[\frac{\d }{\d T}\frac{\kappa T}{1+\kappa}\right]\z{\partial_t + \bvec v \bvec \nabla} T +
\bvec \nabla \bvec v & = & 0,   \label{e:T-class2} \\
 T \sigma (\partial_t  +\bvec v \bvec\nabla) \bvec v &=& - \nabla p , \label{e:Euler-class2} \\
(\kappa + 1)p & = & T\sigma .
\end{eqnarray}

These systems of partial differential equations, together with the formal similarity between the continuity equation,
eq.~(\ref{e:cont}) and the entropy equation, eq.~(\ref{e:entropy}) indicate that Cases 1 and 2 can be solved similarly,
but the time evolution of the temperature will be different in both cases, and there will be other interesting differences.

We search for parametric solutions of these partial differential equations, using in both cases
the same  ansatz for the velocity field $\bvec v$ and the same set of scaling variables, that depend
parametrically on time. Thus our goal is to reduce the system of partial differential equations to
a set of ordinary differential equations that describe the time evolution of the scale parameters. With
this step, the hydrodynamical equations are considered to be solved as the solution of ordinary differential
equations is a built-in routine to several mathematical software packages. In addition, we also relate
these ordinary differential equations to a motion of a mass-point in a classical potential and provide
some first integrals of this motion.

\section{Ansatz of a rotating flow}
We assume that the expanding fireball will rotate and expand simultaneously. The mentioned exact non-relativistic solutions
of refs.~\cite{Csorgo:1998yk,Csorgo:2001ru,Csorgo:2001xm}, where ellipsoidally symmetric, directional Hubble flows
were found, suggest that such Hubble flows are a key component of the velocity field. We assume that the rotation
is characterized by a time-dependent angular velocity, so our ansatz is the following:
\begin{eqnarray}
 \bvec v & =&  \bvec v_H + \bvec v_R,  \label{e:vsol} \\
 \bvec v_H & = & (\frac{\dot X}{X} r_x, \frac{\dot Y}{Y} r_y, \frac{\dot Z}{Z} r_z), \label{e:vH} \\
 \bvec v_R & = & \bvec \omega \times \bvec r \, = \, (- \omega r_y, \omega r_x, 0) ,\label{e:vR}
\end{eqnarray}
so we assume that the principal axes of an expanding ellipsoid are given by the time dependent functions
$(X,Y,Z)\equiv (X(t), Y(t), Z(t) ) $ and the rotation is around the $r_z$ axis with a time dependent
angular velocity $\omega \equiv \omega(t)$.

The angular velocity shows up also as the so called {\it vorticity}
of the flow: the vorticity field is defined by ${\bvec \omega}({\bvec r},t) = {\bvec \nabla }\times {\bvec v}$.
It is straigthforward to see that for this ansatz (and thus for all of the solutions described in this manuscript)
the vorticity vector is parallel to the $r_z$ axis, the axis of rotation, and its value depends only on time:
$\bvec \omega({\bvec r},t)  = (0, 0, 2 \omega(t))$.

In this manuscript, we describe spheroidally symmetric solutions of fireball hydrodynamics,
that are characterized by the equality of the principal axes in the plane transverse to the axis of rotation,
so it is natural to introduce the transverse radius parameter as $R \equiv R(t) = X(t) = Y(t)$.
To describe the new exact solutions, the concept of comoving derivative $\cal D$ and the corresponding
scaling variables $s_R$ and $s_Z$ will be utilized:
\begin{eqnarray}
 {\cal D} & \equiv & \partial_t + (\bvec v \bvec \nabla) ,  \label{e:D} \\
 s_R & = & \frac{r_x^2+r_y^2}{R^2}, \label{e:sr} \\
 s_Z & = & \frac{r_z^2}{Z^2}.\label{e:sz}
\end{eqnarray}
It is easy to verify that for spheroidally symmetric flows with $X = Y = R$,
and for the velocity field given above, both $s_R$ and $s_Z$ are good
scaling variables, that is, their comoving derivative vanishes independently
of the actual time dependence of the scale parameters $(X,Y,Z)$ and $\omega$:
\begin{equation}
{\cal D} s_R = {\cal D} s_Z = 0. 
\end{equation}
We assume that the thermodynamical fields like the temperature $T$, the pressure $p$, the entropy density
$\sigma$ and/or the number density $n$ depend on the coordinates only through some so-called \emph{scaling functions},
that is, functions of the scaling variables $s_R$ and $s_Z$.

Before going on to the discussion of the new solutions, it is interesting to note another appearance of these
scaling variables. For our ansatz for the velocity field, the trajectories of the fluid elements
can easily be determined as follows: denoting the coordinate vector of a fluid element by $\bvec r(t)$, its
equation of motion can be written as
\begin{equation}
\dot{\bvec r}(t) = \bvec v(\bvec r(t), t) .
\end{equation}
Using eqs.(\ref{e:vsol}), (\ref{e:vH}), and (\ref{e:vR}), the solution to this equation is most easily written in
cylindrical coordinates, where these coordinates  of the $\bvec r(t)$ trajectories are denoted by $\rho(t)$, $\varphi(t)$, $z(t)$:
\begin{eqnarray}  
\rho(t) & =&  \rho_0\frac{R(t)}{R(t_0)} , \label{e:tr-rho} \\
\varphi(t) & =&  \varphi_0 + \int \d t \, \omega(t) , \label{e:tr-phi} \\
z(t) & =& z_0 \frac{Z(t)}{Z(t_0)} . \label{e:tr-z}
\end{eqnarray}
These equations show that the trajectories rotate and expand together with the fluid: they follow the time evolution of the
$s_R$ and $s_Z$ scaling variables in the radial and axial directions, while the angular velocity of the trajectory is that of
the fluid, respectively.

\section{New exact, rotating solutions}
The form of the solutions for both {\it Case 1} and {\it Case 2} depends substantially on the form
of the equations of state. If the ratio of the energy density to the pressure is independent of the temperature, 
or $\kappa(T) = const$, a new family of solutions is obtained in both cases, called {\it Case 1A} and {\it Case 2A},
where the scaling function for the temperature profile can be chosen as an arbitrary positive definite function,
while the number density or the entropy density scaling functions are determined by the scaling function
for the temperature. Given this functional degree of freedom, the number of new solutions is infinite in
both {\it Case 1A} and {\it Case 2A}. However, if the ratio $\kappa(T) = \varepsilon/p$ becomes a truly temperature
dependent function, we obtain significantly stronger constraints on the possible form of the temperature
profile, and solutions are obtained only in case of spatially homogeneous but time dependent temperature
profiles. These cases will be called {\it Case 1B} and {\it Case 2B}\footnote{
In both cases, there is a 3rd possibility: for  special forms of the $\kappa(T)$ function, solutions
with arbitrary temperature profiles can be obtained. However, considering possible applications of our solutions
in heavy ion physics, these 
special forms of the $\kappa(T)$ functions are not feasible hence not detailed. These classes 1C and 2C 
can be obtained from the criteria of  the separability of 
the coordinate and the time dependence in the temperature equation, which gives a (solvable) differential
equation for the form of the $\kappa(T)$ function.
}.

As mentioned before, the solutions detailed below correspond to spheroidally symmetric
expansions, that is, $X(t)=Y(t)\equiv R(t)$, but $Z(t)\neq R(t)$ in general. In all the cases
described below, the scaling variable has the same form, 
\begin{equation}
s = s_R + s_Z, 
\end{equation}
and the time dependence of the angular velocity
is also of the same form in terms of the transverse radius $R$. The velocity field follows our
ansatz detailed above, but in each cases a different system of ordinary differential
equations govern the time evolution of the scale parameters $R$ and $Z$.
The common features of the all the solutions given in this paper are
summarized as follows:
\begin{eqnarray}
 \bvec v & =&  \bvec v_H + \bvec v_R, \label{e:vsolsol} \\
 \bvec v_H & = & (\frac{\dot X}{X} r_x, \frac{\dot Y}{Y} r_y, \frac{\dot Z}{Z}
r_z),  \label{e:vHsol} \\
 \bvec v_R & = & \bvec \omega \times \bvec r \, = \, (- \omega r_y, \omega r_x,
0), \label{e:vRsol} \\
 \omega & = & \omega_0 \frac{R_0^2}{R^2} ,\label{e:omega}\\
 R  & = & X = Y \ne Z, \\
 s & = & s_R + s_Z = \frac{r_x^2 + r_y^2}{R^2} +\frac{r_Z^2}{Z^2},\\
 V & = & X Y Z = R^2 Z .
\end{eqnarray}

{\it Case 1A}: Solutions for the case of a conserved charge and a constant,
temperature independent $\kappa$ are given as follows: 
\begin{eqnarray}
 n & = & n_0 \frac{V_0 }{V} {\cal \nu}(s) , \label{e:class1A_nsol}\\
 T & = & T_0 \left(\frac{V_0 }{V} \right)^{1/\kappa} {\cal T}(s) , \label{e:class1a_Tsol}\\
 {\cal\nu}(s) & = & \frac{1}{{\cal T}(s)} \exp\left(- \frac{1}{2}\int_0^s \frac{{\mathrm d}u}{{\cal T}(u)} \right), \\
 R \ddot R - R^2 \omega^2 & = & Z \ddot Z \, = \,  \frac{T_0}{m} \left(\frac{V_0
}{V} \right)^{1/\kappa}, \label{e:class1-sol}
\end{eqnarray}
where the following constants are introduced: $n_0$, the value of the conserved charge density at the origin at
initial time $t_0$, while the initial values of the scale parameters are given by $(X_0, Y_0, Z_0)$.
(Note that we also introduced already at this step a suitable re-scaling, which corresponds to
re-defining the lenght-scales and the temperature scales in terms of the constants of the integrations so
that the solutions have the most transparent forms.) 
The initial proper volume  is $ V_0  =  X_0 Y_0 Z_0 = R^2_0 Z_0 $, and the
scaling function for the temperature and the density profile are normalized so
that ${\cal \nu}(0) = {\cal T}(0) = 1.$ In this class of solutions, the shape
of the temperature profile function, ${\cal T}(s)$ is an arbitrary properly
normalized positive function, so this corresponds to an uncountably infinite number
of solutions,  that follows the lines of refs.~\cite{Csorgo:1998yk,Csorgo:2001ru}.
The time evolution of the scale parameters is given by their initial values,
$X_0 = Y_0 = R_0$ and $ Z_0$, and initial expansion rates 
$\dot X_0 = \dot Y_0 = \dot R_0$ and $\dot Z_0$, together with the dynamical
equations of eq.~(\ref{e:class1-sol}). 
It is beautiful to observe that for the case of vanishing initial angular
momentum, $\omega_0 = 0$, we recover the spheroidal ($X = Y = R$) limit of the
general class of ellipsoidally symmetric solutions that were described in
ref.~\cite{Csorgo:2001ru}.

{\it Case 1B}: These solutions are valid for the case of conserved charge and a
non-constant, temperature dependent $\kappa(T)$ function. These solutions are
written as: 
\begin{eqnarray}
 n & = & n_0 \frac{V_0 }{V} \exp(-s/2) , \\
 T & \equiv & T(t)  , \\
 R \ddot R - R^2 \omega^2 & = & Z \ddot Z \, = \,  \frac{T}{m} 
\label{e:class1b-sol},\\
\frac{d(\kappa T)}{dT}\frac{\dot T}{T} + \frac{\dot V}{V} & =& 0. \label{e:class1b_T}
\end{eqnarray}
where the temperature $T$ is now spatially homogeneous but depends on time,
and the density profile corresponds to an expanding spheroidal Gaussian profile.
The equation for the time evolution of the temperature is exactly the 
same as eq. (9) of ref.~\cite{Csorgo:2001xm}, indicating that the rotation
of the fluid enters the system of the hydrodynamical equations in a rather
straightforward manner. The equation of the time evolution of the temperature
can be solved as follows: we introduce the $f_{1B}(T)$ function as
\begin{equation}
f_{1B}(T) \equiv \exp\left\{\int_{T_0}^T\frac{dy}{y}\left(\frac{d}{dy}\left[\kappa(y)y\right]\right)\right\} ,
\label{e:f1Bdef}
\end{equation}
and with this we can rewrite eq.~(\ref{e:class1b_T}) as follows:
\begin{equation}
\frac{\dot f_{1B}(T)}{f_{1B}(T)} + \frac{\dot V}{V} = 0 \quad\Leftrightarrow\quad
\frac{d}{dt}\left(V f_{1B}(T)\right) = 0 .
\label{e:T1Brewrite}
\end{equation}
If the above $f_{1B}$ function is injective (ie. it has an inverse), then we can readily express the
time dependence of the temperature as
\begin{equation}
T(t) = f^{-1}_{1B}\left(\frac{V_0}{V(t)}\right) .
\label{e:T1Bsol}
\end{equation}
The condition that the $f_{1B}$ function has an inverse is equivalent to requiring $\frac{d}{dT}(\kappa T) > 0$ for
all $T$. There can be special cases of $\kappa(T)$ functions where this condition is not met: in this cases the
temperature is not uniquely determined by the volume. (For an example, we refer to the discussion in
Ref.~\cite{Csanad:2012hr}). Nevertheless, eq.~(\ref{e:T1Brewrite}) is always a valid
equation for the time dependence of the temperature in this case.

It is interesting to see that the solution corresponding to the
case of a temperature independent $\kappa$ and a spatially homogeneous temperature profile
emerges as a common special case of both the 1A and 1B cases. One can verify that for
temperature independent $\kappa(T) = \kappa = const$, eq.~(\ref{e:T1Bsol}) indeed yields
the time evolution of the temperature as given in eq.~(\ref{e:class1a_Tsol}). 
It is also beautiful to observe that for the case of vanishing initial
angular momentum, $\omega_0 = 0$, we recover the spheroidal ($X = Y = R$) limit
of the ellipsoidally symmetric exact Gaussian solutions of non-relativistic
hydrodynamics, that were described for the same, rather general class of
temperature dependent $\kappa\equiv \kappa(T)$ equations of state in
ref.~\cite{Csorgo:2001xm}. Thus we have successfully generalized another
class of exact solutions to the case of rotating and expanding, spheroidal
fireballs. It is also worthwhile to mention, that the evolution equation for
the temperature profile, that was found first in ref.~\cite{Csorgo:2001xm},
is form invariant even if the kinematics is generalized to the relativistic
region following the exact relativistic hydro solutions described in
ref.~\cite{Csanad:2012hr}: equation (9) of ref.~\cite{Csorgo:2001xm} and
equation (16) of ref.~\cite{Csanad:2012hr} are equivalent, if the
relativistic co-moving derivative in ref.~\cite{Csanad:2012hr} is evaluated
in the non-relativistic limit, where it corresponds to
the time derivative.

{\it Case 2A}: This case corresponds to local conservation of entropy, with a temperature
independent $\kappa$. The solutions are similar also to the solutions in {\it Case 1A},
but now the entropy density takes the role of the density of conserved charge:
it has also an arbitrary scaling function ${\cal S}(s)$, but the relation
between the scaling functions of the temperature and the entropy density becomes
modified as compared to the relations in {\it Case 1A}. The solution is
written as follows:
\begin{eqnarray}
 \sigma & = & \sigma_0 \frac{V_0 }{V} {\cal S}(s) , \label{e:class2A_ssol}\\
 T & = & T_0 \left(\frac{V_0 }{V} \right)^{1/\kappa} {\cal T}(s) , \label{e:class2a_Tsol}\\
 {\cal S}(s) & = & \frac{1}{{\cal T}(s)} \exp\left(- s/2 \right), \\
 R \ddot R - R^2 \omega^2 & = & Z \ddot Z \, = \, \frac{1}{1+\kappa}.
\label{e:class2a-sol}
\end{eqnarray}
It is important to note, that this case also corresponds to finite, expanding
fireball, given that the pressure $p = \sigma T/(1 + \kappa)$ decreases as
a Gaussian in coordinate space for any choice of the scaling function for the
temperature. Note also that in {\it Case 2A} the explosion of the fireball is
significantly faster, than in case of {\it Case 1A}: in {\it Case 1A} the acceleration is
driven by of $ p/(nm) = nT/(nm) = T/m$ that tends to vanishing values
at late times, given that the fireball expands hence the temperature decreases.
In contrast, in the current case of {\it Case 2A} where only entropy is conserved
locally, but there is no conserved charge, the acceleration in the Euler
equation is driven by $p/(\varepsilon +p) = 1/(1 + \kappa)$ which is a
constant. If such a system approaches a soft point, where $\kappa \rightarrow
\infty$, the accelerations due to pressure gradients tend to vanishing values.
Nevertheless, acceleration due to non-vanishing initial angular velocity will
still be present in the plane perpendicular to the axis of rotation.

{\it Case 2B} solutions correspond also to the case when there is no
conserved charge but the expansion is adiabatic and the thermodynamical
relations are characterized by a non-constant, temperature dependent
$\kappa \equiv \kappa(T)$ function. These solutions can be written as follows: 
\begin{eqnarray}
 \sigma & = & \sigma_0 \frac{V_0 }{V} \exp(-s/2) , \\
 T & \equiv & T(t)  , \\
 R \ddot R - R^2 \omega^2 & = & Z \ddot Z \, = \,  \frac{1}{1+\kappa(T)} 
\label{e:class2b-sol},\\
\left[\frac{T}{1+\kappa}\frac{d\kappa}{dT} +\kappa\right]\frac{\dot T}{T}
+ \frac{\dot V}{V} & = & 0. \label{e:class2b_T}
\end{eqnarray}
where the temperature $T$ is spatially homogeneous but depends on time,
similarly to the {\it Case 1B} solutions, but with a modified evolution equation,
and the entropy density profile corresponds to an expanding Gaussian spheroid.
It is interesting to see that the equation for the time evolution of the
temperature is exactly the same as eq. (17) of the relativistic hydro solution of
ref.~\cite{Csanad:2012hr}. The equation for the time evolution of the
temperature can be solved in a manner similar to that of {\it Case 1B} above:
we introduce the $f_{2B}(T)$ function as
\begin{equation}
f_{2B}(T) \equiv
\exp\left\{\int_{T_0}^Tdy\left(\frac{\kappa(y)}{y}+\frac{\frac{d\kappa(y)}{dy}}{\kappa(y)+1}\right)\right\} ,
\label{e:f2Bdef}
\end{equation}
so that we can rewrite eq.~(\ref{e:class2b_T}) as follows:
\begin{equation}
\frac{\dot f_{2B}(T)}{f_{2B}(T)} + \frac{\dot V}{V} = 0 \quad\Leftrightarrow\quad
\frac{d}{dt}\left(Vf_{2B}(T)\right) = 0 ,
\label{e:T2Brewrite}
\end{equation}
which again can be solved if the $f_{2B}$ function has an inverse. The condition on the equation
of state which states that $f_{2B}$ should have an inverse is in this case (as can easily be shown)
equivalent to $\frac{d}{dT}\frac{T\kappa(T)}{1+\kappa(T)}>0$, which is again equivalent to
$\frac{d}{dT}\frac{\varepsilon}{\sigma} > 0$. To our best knowledge, this condition is met for
any reasonable, thermodynamically consistent equation of state.

If the $f_{2B}$ function has an inverse, then we can express the time dependence of the temperature as
\begin{equation}
T(t) = f^{-1}_{2B}\left(\frac{V_0}{V(t)}\right) .
\label{e:T2Bsol}
\end{equation}
Again, if $\kappa(T) = \kappa = const$, we recover the time evolution of the temperature
as in {\it Case 2A}, which is given by eq.~(\ref{e:class2a_Tsol}). 

We think that {\it Case 2B} is an important one, given that
the equations of state under consideration include, as a particular
realization, the lattice QCD equations of state, as was demonstrated in ref.
~\cite{Csanad:2012hr} and references therein. This result improves in two ways
on two shortcomings of ref.~\cite{Csanad:2012hr}: First, the 
acceleration of the fluid, that was assumed to be, in the relativistic sense, a
non-accelerating Hubble flow, can now be considered explicitly, but at present 
only in the non-relativistic kinematic domain. Second, the initial conditions
in high energy heavy ion physics create fireballs with non-vanishing total
angular momentum. Now, the effects of the initial angular momentum can be
evaluated explicitly, using $\omega_0 \ne 0$ initial conditions.
However, the detailed evaluation of the observables from these class of solutions
goes beyond the scope of the current manuscript.

Note that the equations for the temperature can be integrated both in the
{\it Case 1B} and {\it Case 2B} solutions, similarly to the results of 
refs.~\cite{Csorgo:2001xm,Csanad:2012hr}, respectively.

It took 35 years~\cite{nagymarciPhD,csorgo:zimanyi2012} after the 
publication of the first exact hydrodynamical solutions in a similar class~\cite{Bondorf:1978kz},
to find and write up the {\it rotating} non-relativistic solutions of perfect fluid hydrodynamics.
However, once they are given, it is straightforward to check their validity. 

\section{First integrals of the classical motion}
The total angular momentum $\mathbf{M}$ of the fluid is constant. This quantity is defined as
\[
\mathbf{M} = \int dr_x dr_y dr_z\,mn\mathbf{r}\times\mathbf{v}
\]
in the case when there is a conserved particle number present (Cases 1A and 1B).
For the solutions discussed in this manuscript here one immediately obtains that 
only the $z$ component of $\mathbf{M}$ (denoted by $M_z$) is non-vanishing (which is
clear also from the shape of the velocity field). One can use eq.~(\ref{e:class1A_nsol})
for the expression of $n$ in Case 1A, and also for Case 1B (which from this point of
view is a special case of Case 1A, with $\nu(s)=e^{-s/2}$). With a re-scaling of the
integration variables and transforming the integral to a
$r$, $\varphi$, $z$ cylindrical coordinate system (ie. $r_x\to Ar\cos\varphi$,
$r_y\to Br\sin\varphi$, $r_z\to Bz$), one gets the following expression in Cases
1A and 1B:
\begin{equation}
M_z = C\cdot m\omega R^2,
\end{equation}
where the constant $C$ is introduced as
\begin{equation}
C \equiv 2\pi n_0V_0\int dr dz\,r^2\nu\left(r^2+z^2\right) .
\end{equation}

One readily identifies $CmR^2$ as the moment of inertia of the fluid $\Theta$, where
the constant $C$ encodes the shape of the mass (or entropy) distribution of the fluid.
Thus we have
\begin{equation}
 M_z \equiv \Theta \omega \propto m R^2 \omega_0 \frac{R_0^2}{R^2} = m \omega_0 R_0^2 .
\end{equation}
So eq.~(\ref{e:omega}), ie. the relation $\omega R^2 = \omega_0 R_0^2$, which is true
for all the solutions discussed in this paper, is established as a consequence of the conservation
of the total angular momentum of the fluid.

In the other case when there is no conserved $n$ (Cases 2A and 2B), the definition of
$\mathbf{M}$ is not so straightforward. If eg. one applies Landau's argument to say
that in the final state the number of particles produced is proportional to the entropy
density $\sigma$, one can tentatively define $\mathbf{M}$ as $\mathbf{M} \propto\int
dr_x dr_y dr_z\,\sigma\mathbf{r}\times\mathbf{v}$. With this definition, an analogue
of the calculation above leads again to the result that the relation $\omega R^2 =
\omega_0 R_0^2$ is a consequence of angular momentum conservation. However, the discussion
whether this definition of the angular momentum indeed should be applied to all cases
with no conserved particle number goes beyond the scope of this manuscript.

Investigating the equations of motion for the scale parameters leads to other interesting
observations regarding conserved quantities. One can show that the equations of motion for
the scale parameters in the discussed solutions are equivalent to the Lagrangian equations
of motion of a point mass moving in a non-central potential $U$.
The presence of non-vanishing angular momentum yields an extra term
in the potential as compared to the same Hamiltonian motion, introduced and
discussed in ref.~\cite{Csorgo:2001ru}: the coordinates are $(X, Y,Z)$,
the momentum components are $(P_x, P_y, P_z) = m (\dot X, \dot Y, \dot Z)$.
The initial conditions reflect the fact that our solutions are spheroidally symmetric:
\begin{eqnarray}
 X_0 & = & Y_0 \, = \,  R_0, \\
 \dot X_0 & = & \dot Y_0 = \dot R_0.
\end{eqnarray}
The Hamiltonian of the motion for the {\it Case 1A} solutions is
\begin{eqnarray}
 H_{1A} & =&  \frac{P_x^2 + P_y^2 + P_z^2}{2m} + U_{1A}, \\
U_{1A} & =& \kappa T_0 \left(\frac{X_0 Y_0 Z_0}{X Y Z}\right)^{1/\kappa} +
\frac{2m \omega_0^2 R_0^4}{X^2 + Y^2},
\end{eqnarray}
Using similar notation, the equation of motion for the solutions in {\it Case 2A}
can also be derived from a classial Hamiltonian, with a modified, logarithmic
potential:
\begin{eqnarray}
 H_{2A} & =&  \frac{P_x^2 + P_y^2 + P_z^2}{2m} + U_{2A}, \\
U_{2A} & = & - \frac{m }{1 + \kappa} \ln \left(\frac{X Y Z}{X_0 Y_0 Z_0}\right) +
\frac{2m \omega_0^2 R_0^4}{X^2 + Y^2}.
\end{eqnarray}
More generally, using the solutions for the time dependence of the temperature in Cases 1B and 2B,
(eqs.~(\ref{e:f1Bdef})--(\ref{e:T1Bsol}) for {\it Case 1B} and eqs.~(\ref{e:f2Bdef})--(\ref{e:T2Bsol}) for {\it Case 2B}),
the equations of motion for the {\it Case 1B} and {\it Case 2B} solutions can also be cast in a Hamiltonian form.
For this purpose, we introduce the following functions of the volume $V \equiv XYZ$:
\begin{eqnarray}
\chi_{1B}(V) \equiv & \int_{V_0}^V\frac{dV'}{V'}T ,            \quad & T \equiv T(V') \quad(1B) , \\
\chi_{2B}(V) \equiv & m\int_{V_0}^V\frac{dV'/V'}{1+\kappa(T)} , \quad & T \equiv T(V') \quad(2B) , 
\end{eqnarray}
where we indicated that the temperature is expressed as a function of the volume $V'$, using eq.~(\ref{e:T1Bsol}) or
eq.~(\ref{e:T2Bsol}) for Cases 1B and 2B, respectively. With these definitions, one can verify that the equations of motion
for {\it Case 1B} stem from the following Hamiltonian:
\begin{eqnarray}
 H_{1B} & =&  \frac{P_x^2 + P_y^2 + P_z^2}{2m} + U_{1B}, \\
U_{1B} & = & - \chi_{1B}(XYZ) + \frac{2m \omega_0^2 R_0^4}{X^2 + Y^2},
\end{eqnarray}
and for {\it Case 2B} we have
\begin{eqnarray}
 H_{2B} & =&  \frac{P_x^2 + P_y^2 + P_z^2}{2m} + U_{2B}, \\
U_{2B} & = & - \chi_{2B}(XYZ) + \frac{2m \omega_0^2 R_0^4}{X^2 + Y^2}.
\end{eqnarray}
From the clear Hamiltonian structure of the motion of the mass-point,
first integrals of the motion can be found easily by requiring that the total
energy of the mass point during its classical, conservative motion is conserved.
These first integrals can also be parameterized, similarly to the case
discussed in ref.~\cite{Akkelin:2000ex}, by the asymptotic values of the
expansion velocities of the scales $(\dot X_a, \dot Y_a, \dot Z_a)$, since for late
times the acceleration vanishes in all the considered classes of exact solutions.
These first integrals are thus given as
\begin{eqnarray}
 H_{1A} & = & E_{1A}\, = \, const \, = \, \frac{1}{2} m (\dot X_a^2 + \dot Y_a^2 + \dot Z_a^2), \\
  H_{2A} & = & E_{2A} \, = \, const \, =\, \frac{1}{2} m (\dot X_a^2 + \dot Y_a^2 + \dot Z_a^2).
\end{eqnarray}
Formally similar equations hold for Cases 1B and 2B.

It is also interesting to remark that for a spheroid that rotates along its $Z$
axis with radius in the plane of rotation denoted by $R$ the moment of inertia 
$\Theta$ is proportional to $R^2$ and the kinetic energy of the rotation $E_R$ is 
\begin{equation}
 E_R \propto \Theta \omega^2 \propto R^2 \omega^2.
\end{equation}
Given that in our case the time evolution of the angular velocity $\omega$
is expressed as $\omega_0 R_0^2 = \omega R^2$,
one can also easily identify the last term of the Hamiltonians given above
as the kinetic energy of the rotation,
\begin{equation}
 E_R = \frac{2m \omega_0^2 R_0^4}{X^2 + Y^2} = m R^2 \omega^2 .
\end{equation}
This term is obviously the same for all the discussed cases, $1A$, $1B$, $2A$, $2B$.
Given that for late times $X= Y = R$ tends to infinity, one can also
see that although the the total energy is conserved, the kinetic energy carried 
by the rotation tends to vanishing values for late times. This is also 
a rather natural consequence of the expansion and the related increase of
the moment of inertia, which thus stops the rotation of the fluid at asymptotically
late times.

Using the solution of these classical Hamiltonian problems, the time evolution
of the fluid trajectories is given by the solution of the trajectory equation
$\dot {\bvec r}(t) = {\bvec v}({\bvec r}(t),t)$, which was given by 
eqs.~(\ref{e:tr-rho},\ref{e:tr-phi},\ref{e:tr-z}) in cylindrical coordinates, but for
clarity let us rewrite them in Cartesian coordinates of
the trajectories ${\bvec r}(t) = (x(t), y(t), z(t))$ in the following compact form:
\begin{eqnarray}
 x(t) & = & r_0 \frac{\omega_0}{\omega} \cos\left[\phi_0 + \int_{t_0}^t \mathrm{d}t \omega(t) \right], \label{e:tr-x} \\
 y(t) & = & r_0 \frac{\omega_0}{\omega} \sin\left[\phi_0 + \int_{t_0}^t \mathrm{d}t \omega(t) \right], \label{e:tr-y} \\
 z(t) & = & z_0 \frac{Z}{Z_0} , \label{e:tr-z2} 
\end{eqnarray}
where ${\bvec r}_0 = (x_0, y_0, z_0) = (r_0 \cos(\phi_0), r_0 \sin(\phi_0), z_0)$ denote the initial conditions
for a given fluid-trajectory.

\section{Summary and outlook}

We described and analyzed a class of exact solutions for rotating,
exploding fireball hydrodynamics in four different cases of
the equation of state. We have generalized the equations of non-relativistic hydrodynamics
to the case of vanishing conserved charge, a scenario relevant for the lattice QCD equation
of state in the vanishing net baryon density or $\mu_B = 0$ limit. We have found that in this
case (discussed in the manuscript as Cases 2A and 2B) the expansion is more explosive than in
Cases 1A and 1B, when the equation of state corresponds to locally conserved number of particles.
This feature follows not only from the analytic form of the dynamical equations, but also can
readily be seen from the explicitly given forms of the solutions of the equations for the fluid
trajectories. One can see that a fluid without a conserved charge and a temperature dependent
speed of sound expands faster radially hence slows down its rotation faster than a similar
fluid with a conserved charge, whose pressure decreases relatively faster.

In the limit of vanishing angular momentum, our solutions
simplify to earlier, irrotational exact solutions of non-relativistic fireball hydrodynamics.
Our results show similarity to the first exact solution of rotating and expanding, relativistic
fireballs~\cite{Nagy:2009eq,nagymarciPhD}, 
and also to recent numerical solutions of hydrodynamics for rotating and exploding fireballs~\cite{Csernai:2013uda,Csernai:2013vda,Csernai:2013add}.

The equations of motion of the scale variables can be cast in a form that corresponds
to a classical motion of a massive particle in a conservative potential, described by appropriate
Hamiltonians. From the  analysis of this Hamiltonian motion, we have found two first integrals
of the hydrodynamical problem, that correspond to the conservation of the total 
energy $E$ and the total angular momentum $M_z$. We have expressed the total energy
in terms of the asymptotic expansion rates of the fluid in the principal directions
and the total angular momentum in terms of the initial transverse size and the initial angular
velocity of the rotating fluid.

Further work, in particular the evaluation of the effects of rotation on the observable
quantities is necessary to make a comparison with existing data or observables obtained from
numerical hydrodynamical solutions, and to suggest new,  experimentally well measurable 
observables that characterize the rotation.
These questions go beyond the scope of the present manuscript, except the obvious
comment that the total angular momentum of the fireballs created in high energy
heavy ion collisions should be measured experimentally as it is directly related
to the initial transverse radius and the initial angular velocity of the fluid.

\section{Acknowledgments}
This research has been supported by a Ch. Simonyi Fellowship and by  
the Hungarian  OTKA NK 101438 grant.


\begin{thebibliography}{99}
\bibitem{Csernai:2013uda} 
  L.~P.~Csernai and S.~Velle,
  arXiv:1305.0385 [nucl-th].

\bibitem{Csernai:2013vda} 
  L.~P.~Csernai, S.~Velle and D.~J.~Wang,
  arXiv:1305.0396 [nucl-th].

\bibitem{Csernai:2013add} 
  F. Becattini, L.~P.~Csernai and D.~J.~Wang,
  arXiv:1304.4427 [nucl-th].

\bibitem{Bondorf:1978kz} 
  J.~P.~Bondorf, S.~I.~A.~Garpman and J.~Zim\'anyi,
  Nucl.\ Phys.\ A {\bf 296}, 320 (1978).
  
\bibitem{Csizmadia:1998ef} 
  P.~Csizmadia, T.~Cs\"org\H{o} and B.~Luk\'acs,
  Phys.\ Lett.\ B {\bf 443}, 21 (1998)
  [nucl-th/9805006].

\bibitem{Csorgo:1998yk} 
  T.~Cs\"org\H{o},
  Central Eur.\ J.\ Phys.\  {\bf 2}, 556 (2004)
  [nucl-th/9809011].

\bibitem{Akkelin:2000ex} 
  S.~V.~Akkelin, T.~Cs\"org\H{o}, B.~Luk\'acs, Y.~.M.~Sinyukov and M.~Weiner,
  Phys.\ Lett.\ B {\bf 505}, 64 (2001)
  [hep-ph/0012127].

\bibitem{Csorgo:2001ru} 
  T.~Cs\"org\H{o},
  Acta Phys.\ Polon.\ B {\bf 37}, 483 (2006)
  [hep-ph/0111139].
  
\bibitem{Csorgo:2001xm} 
  T.~Cs\"org\H{o}, S.~V.~Akkelin, Y.~Hama, B.~Luk\'acs and Y.~.M.~Sinyukov,
  Phys.\ Rev.\ C {\bf 67}, 034904 (2003)
  [hep-ph/0108067].

\bibitem{Landau:1953gs} 
  L.~D.~Landau,
  Izv.\ Akad.\ Nauk Ser.\ Fiz.\  {\bf 17}, 51 (1953).

\bibitem{Belenkij:1956cd} 
  S.~Z.~Belenkij and L.~D.~Landau,
  Nuovo Cim.\ Suppl.\  {\bf 3S10}, 15 (1956)
  [Usp.\ Fiz.\ Nauk {\bf 56}, 309 (1955)].



\bibitem{Hwa:1974gn} 
  R.~C.~Hwa,
  Phys.\ Rev.\ D {\bf 10}, 2260 (1974).

\bibitem{Bjorken:1982qr} 
  J.~D.~Bjorken,
  Phys.\ Rev.\ D {\bf 27}, 140 (1983).
  
\bibitem{Csorgo:2002ki} 
  T.~Cs\"org\H{o}, F.~Grassi, Y.~Hama and T.~Kodama,
  Heavy Ion Phys.\ A {\bf 21}, 53 (2004)
  [Acta Phys.\ Hung.\ A {\bf 21}, 53 (2004)]
  [hep-ph/0203204].

\bibitem{Csorgo:2002bi} 
  T.~Cs\"org\H{o}, F.~Grassi, Y.~Hama and T.~Kodama,
  Heavy Ion Phys.\ A {\bf 21}, 63 (2004)
  [Acta Phys.\ Hung.\ A {\bf 21}, 63 (2004)]
  [hep-ph/0204300].
  
\bibitem{Csorgo:2003rt} 
  T.~Cs\"org\H{o}, F.~Grassi, Y.~Hama and T.~Kodama,
  Phys.\ Lett.\ B {\bf 565}, 107 (2003)
  [nucl-th/0305059].
  
\bibitem{Csorgo:2003ry} 
  T.~Cs\"org\H{o}, L.~P.~Csernai, Y.~Hama and T.~Kodama,
  Heavy Ion Phys.\ A {\bf 21}, 73 (2004)
  [nucl-th/0306004].

  
\bibitem{Csorgo:2008pe} 
  T.~Cs\"org\H{o}, M.~I.~Nagy and M.~Csan\'ad,
  J.\ Phys.\ G {\bf 35}, 104128 (2008)
  [arXiv:0805.1562 [nucl-th]].

\bibitem{Csorgo:2006ax} 
  T.~Cs\"org\H{o}, M.~I.~Nagy and M.~Csan\'ad,
  Phys.\ Lett.\ B {\bf 663}, 306 (2008)
  [nucl-th/0605070].

\bibitem{Nagy:2007xn} 
  M.~I.~Nagy, T.~Cs\"org\H{o} and M.~Csan\'ad,
  Phys.\ Rev.\ C {\bf 77}, 024908 (2008)
  [arXiv:0709.3677 [nucl-th]].


\bibitem{Nagy:2009eq} 
  M.~I.~Nagy,
  Phys.\ Rev.\ C {\bf 83}, 054901 (2011)
  [arXiv:0909.4285 [nucl-th]].
   
\bibitem{nagymarciPhD}
       M. I. Nagy, PhD Thesis, ELTE University, 2012 (in Hungarian),\\
       \texttt{www.doktori.hu/index.php?menuid=193{\&}vid=10335} 

\bibitem{Wong:2008ex} 
  C.~Y.~Wong,
  Phys.\ Rev.\ C {\bf 78}, 054902 (2008)
  [arXiv:0808.1294 [hep-ph]].

\bibitem{Jiang:2013rm} 
  Z.~J.~Jiang, Q.~G.~Li and H.~L.~Zhang,
  J.\ Phys.\ G {\bf 40}, 025101 (2013).

\bibitem{Zetenyi:2010jp} 
  M.~Zetenyi and L.~P.~Csernai,
  Phys.\ Rev.\ C {\bf 81}, 044908 (2010)
  [arXiv:1003.3757 [nucl-th]].

\bibitem{Tamosiunas:2011qg} 
  K.~Tamosiunas,
  Eur.\ Phys.\ J.\ A {\bf 47}, 121 (2011)
  [arXiv:1106.4839 [hep-ph]].

\bibitem{Bialas:2007iu}
  A.~Bialas, R.~A.~Janik and R.~B.~Peschanski,
  Phys.\ Rev.\  C {\bf 76}, 054901 (2007).

\bibitem{Beuf:2008vd} 
  G.~Beuf, R.~Peschanski, E.~N.~Saridakis,
  Phys.\ Rev.\ C {\bf 78}, 064909 (2008).

\bibitem{Peschanski:2010cs} 
  R.~Peschanski and E.~N.~Saridakis,
  Nucl.\ Phys.\ A {\bf 849}, 147 (2011)
  [arXiv:1006.1603 [hep-th]].

\bibitem{Liao:2009zg}
  J.~Liao and V.~Koch,
  Phys.\ Rev.\  C {\bf 80}, 034904 (2009). 

\bibitem{Lin:2009kv} 
  S.~Lin and J.~Liao,
  Nucl.\ Phys.\ A {\bf 837}, 195 (2010)
  [arXiv:0909.2284 [nucl-th]].

\bibitem{Gubser:2010ze}
  S.~S.~Gubser,
  Phys.\ Rev.\  D {\bf 82}, 085027 (2010).

\bibitem{csorgo:zimanyi2012}
       T. Cs\"org\H{o}, {\it New rotational solutions of hydrodynamics}
       Talk given at the 2012 Zim\'anyi School on Heavy Ion Physics, Budapest, Hungary, December 4, 2012,
       {\tt http://zimanyischool.kfki.hu/12} .
 
\bibitem{Csanad:2012hr} 
  M.~Csan\'ad, M.~I.~Nagy and S.~L\"ok\"os,
  Eur.\ Phys.\ J.\ A {\bf 48}, 173 (2012)
  [arXiv:1205.5965 [nucl-th]].
  
\end{thebibliography}
\end{document}